\documentclass{PoS}

\PoS{PoS(LAT2005)239}

\usepackage{epsfig}
\usepackage{amsmath,epsfig,amssymb,dsfont}
\usepackage{amscd,latexsym}

\newcommand{\R}{{R_\mathrm{det}}}

\def\lsi{\raise0.3ex\hbox{$<$\kern-0.75em\raise-1.1ex\hbox{$\sim$}}}
\def\gsi{\raise0.3ex\hbox{$>$\kern-0.75em\raise-1.1ex\hbox{$\sim$}}}

\input macros.sty      % \eqalign, \meqalign
\input eqalign.sty     % \eqalign, \meqalign

\title{Testing Fermion Actions: \\ Scaling in the Schwinger Model }

\ShortTitle{Scaling in the Schwinger Model}

\author{Nils Christian$^1$, Karl Jansen$^1$, Kei-ichi Nagai$^1$,
        \speaker{Beatrix Pollakowski} $^{1,2}$\\
        $^1$ John von Neumann-Institut f\"ur Computing NIC, \\             
  ~\hspace{0.05cm} DESY, Platanenallee 6, D-15738 Zeuthen, Germany\\
	$^2$ Institut f\"ur Physik, Humboldt Universit\"at zu Berlin,\\
  ~\hspace{0.05cm} Newtonstrasse 15, D-12489 Berlin, Germany\\
E-mail: \email{Nils.Christian,Karl.Jansen,Keiichi.Nagai,Beatrix.Pollakowski@desy.de}}

\abstract{
  We test the scaling behaviour of Wilson, hypercube, maximally twisted
  mass and overlap fermion actions in dynamical simulations of the
  2-dimensional massive Schwinger model. We also present possibilities
  to simulate overlap fermions dynamically by replacing the exact
  overlap operator by an approximate version. This approximation is used
  either as only the guidance Hamiltonian, keeping the exact overlap
  operator as the accept/reject Hamiltonian or for both, the guidance
  and accept/reject Hamiltonian in the Hybrid Monte Carlo algorithm. In
  the latter case we test whether the approximation error can be
  corrected for by computing the determinant ratio of the exact and the
  approximate overlap operators stochastically. 

\vspace{1.0cm}
{\large
\texttt{HU-EP-05/62}\\
\texttt{DESY 05-194}}
}
\FullConference{XXIIIrd International Symposium on Lattice Field Theory\\
                 25-30 July 2005\\
                 Trinity College, Dublin, Ireland}

\begin{document}

\section{Introduction}

Determining the scaling properties of physical quantities computed
with different lattice actions as used presently for dynamical
fermion simulations is certainly one of the essential steps to perform
a controlled continuum limit.  
Obviously, performing such scaling tests for lattice QCD in four 
dimensions is
(at least) very demanding. Therefore, we have
chosen the 
2-dimensional, massive Schwinger 
model \cite{Schwinger:1962tp} as a test laboratory   
to address the scaling properties of a number of 
fermion actions for $N_f=2$ flavours of dynamical fermions. 
In particular, we will compare
standard Wilson \cite{Wilson:1974sk}, 
Wilson twisted mass \cite{Frezzotti:2000nk,Frezzotti:2003ni} at maximal twist, 
hypercube \cite{Bietenholz:1999km} 
and overlap fermions \cite{Neuberger:1997fp} in 
their approach to the continuum limit. 

In order to study the scaling behaviour, 
we will fix the scaling variable $z\equiv (m_f\sqrt{\beta})^{2/3}$ 
to $z=0.2,0.4,0.8$
where $m_f$ is the fermion mass in lattice units 
derived from the PCAC relation and $\beta$ is the 
coupling multiplying the Wilson plaquette gauge action used throughout this 
work. 
Denoting by $e$ the physical gauge coupling and by $a$ the lattice 
spacing, $\beta=1/a^2e^2$. 
At each of the values of $z$ given above, we compute the pseudo 
scalar mass 
and follow its behaviour as a function of 
(decreasing) values of the lattice spacing.
Performing finally a continuum limit of our results allows us to compare
to analytical predictions that are available from approximations of the 
massive Schwinger model which cannot be solved exactly. 
For a discussion of the scalar condensate we 
refer to ref.~\cite{schwingerpap}, where also a more detailed account of our work
can be found.

Another important aspect discussed in this contribution is a description 
of several ways to perform overlap simulations dynamically. 
We follow the general
idea to replace the exact overlap operator $D_\mathrm{ov}$ 
by some approximate version 
$D_\mathrm{ov}^\mathrm{approx}$ having an infra-red safe kernel. 
To correct for such an approximation we test to reweight
by the determinant ratio 
$\mathrm{det}[D_\mathrm{ov}]/\mathrm{det}[D_\mathrm{ov}^\mathrm{approx}]$.
We also have tested the possibility to use 
$D_\mathrm{ov}^\mathrm{approx}$ as the guidance Hamiltonian in the molecular
dynamics part of the Hybrid Monte Carlo (HMC) algorithm while keeping 
the exact overlap operator 
$D_\mathrm{ov}$ as the accept/reject Hamiltonian.

\section{Scaling results}

%In this section we briefly describe 
%the physical observables we have used and the scaling results for the four
%fermions actions. 
%We refer to ref.~\cite{schwingerpap} for more details. 
%Besides the original works introducing Wilson, 
%\cite{Wilson:1974sk}, twisted mass \cite{Frezzotti:2000nk,Frezzotti:2003ni}, 
%hypercube \cite{Hasenfratz:1998bb} 
%and overlap \cite{Neuberger:1997fp} fermions, 
%ref.~\cite{schwingerpap} can 
%also be consulted for the precise definitions of the different actions
%and the notation we have employed.  

The 
pion mass $m_\pi$ and the PCAC fermion mass $m_\mathrm{PCAC}$ have 
been computed by standard methods, 
using the exponential decay of the 2-point function of the pseudo scalar 
operator $\mathbb{\tilde P}$ and computing  
the PCAC relation,
respectively. 
In table~\ref{tab:setup} we list the techniques we 
have used for the simulations and to compute 
the correlators numerically for the different fermion actions. 
We just note that for the twisted mass (tm) fermion case we worked at
full twist, realized by the vanishing of the 
PCAC fermion mass calculated from pure Wilson fermions. 
In this case, the operators used for the correlation functions need to 
be appropriately rotated \cite{Frezzotti:2003ni}, of course.
We refer to ref.~\cite{schwingerpap} for the simulation details such as the 
statistics of our runs and the lattice size and 
for the precise definitions of the different actions
and the notation we have employed.  
In this reference we also discuss the finite size 
corrections we have performed for the pion masses when necessary.

%Another remark is that
%for the currents that are needed in the PCAC relation one can use, 
%besides the standard local currents, also the 
%conserved currents following the prescription to 
%derive these currents 
%as given in \cite{Hasenfratz:2002rp}.
%We have computed these conserved currents in particular for the complicated
%case of hypercube fermions. However, since we do not use these currents
%for the overlap fermions, we restrict ourselves to the local currents 
%for this contribution, see however ref.~\cite{schwingerpap}. 

\begin{table}
\begin{center}
\begin{tabular}{|l||l|l|}
  \hline
  & Wilson / tm / hypercube & overlap \\
  \hline
  \hline
  simulations & standard HMC &  determinant reweighting \\
  \hline
  operators & conjugate gradient solver & eigenvalues and eigenvectors \\
             &                          & conjugate gradient solver  \\
  \hline
  currents & local / conserved & local\\
  \hline
pseudo scalar correlator & $\langle\mathbb{\tilde P \tilde P}\rangle$ &  $\langle\mathbb{\tilde P \tilde P}-\mathbb{\tilde S \tilde S}\rangle$\\
  \hline
%  disconnected & stochastic & exact\\
%  \hline
%  scalar condensate & Ward-Takahashi-identity & $\langle \bar\psi (1-\frac{a}{2}D)\psi\rangle$\\
  \hline
\end{tabular}
\caption{The simulation techniques and the methods to compute the observables
for the different actions we have used are listed. 
\label{tab:setup}}
\end{center}
\end{table}

\begin{figure}[tbp]
\centering
\includegraphics[width=13.0cm]{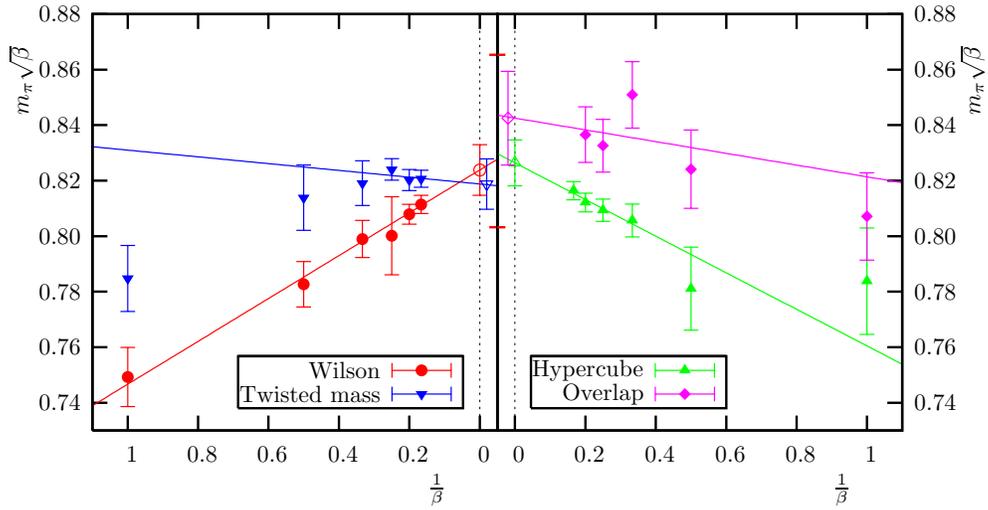}
\caption{Scaling test of $m_\pi\sqrt{\beta}$ as function of 
$1/\beta\propto a^2$ at fixed value of $z=0.4$.}
\label{fig:scaling04}
\end{figure}

In fig.~\ref{fig:scaling04} we show one example of the continuum extrapolation
of $m_\pi\sqrt{\beta}$ for the four different actions we have used. 
We can see that for all the actions there is a nice scaling with $1/\beta$, 
i.e. with $a^2$. Another observation is that for all four actions 
the continuum extrapolated values are consistent, in agreement
with universality. 
Given this finding, we decided to perform 
constraint continuum extrapolations using fits linear in $1/\beta$ with a 
common 
continuum value for all actions at fixed value of $z$. 
These final results from our analysis 
together with theoretical predictions are given in table~\ref{tab:contcomp}.
The theoretical calculations are performed using the Sine-Gordon model 
\cite{Smilga:1996pi}, denoted as SG in the table, and  from a  
large mass expansion 
\cite{Gattringer:1995iq} which we denote as LG. Our own non-perturbatively
obtained values are denoted as NP. As one can see from 
table~\ref{tab:contcomp},
with the high precision we could reach with our simulations, 
deviations from these theoretical expectations exist. Only for a value 
of $z=0.2$ there is a consistency with the prediction of 
ref.~\cite{Smilga:1996pi} while the results of 
ref.~\cite{Gattringer:1995iq} may start to 
describe the data only for $z=0.8$ or larger. 

\begin{table}
\begin{center}
\begin{tabular}{|l||l|l|l|}
\hline
$z=(m_f\sqrt{\beta})^{2/3}$ & $(m_\pi\sqrt{\beta})_\mathrm{SG}$ &
  $(m_\pi\sqrt{\beta})_\mathrm{LG}$ & $(m_\pi\sqrt{\beta})_\mathrm{NP}$ \\
\hline
\hline
0.2 & 0.4016 & 0.43266 & 0.406(4) \\
\hline
0.4 & 0.8032 & 0.86532 & 0.825(5) \\
\hline
0.8 & 1.6064 & 1.73064 & 1.755(4) \\
\hline
\end{tabular}
\caption{Continuum values for $m_\pi\sqrt{\beta}$ at fixed value 
of $z=(m_f\sqrt{\beta})^{2/3}$. We compare our continuum extrapolated
non-perturbatively obtained values, denoted with NP, with theoretical 
predictions using a Sine-Gordon approach (SG, \cite{Smilga:1996pi}) 
and using a large
mass approximation (LG, \cite{Gattringer:1995iq}). 
\label{tab:contcomp}}
\end{center}
\end{table}

\subsection{Simulations}

For the Wilson, the hypercube and the twisted mass
fermion action a standard Hybrid Monte Carlo algorithm (HMC) 
\cite{Duane:1987de} has been implemented for our dynamical simulations.
For overlap fermions, we generated configurations with the pure gauge 
action and then reweighted all observables with the overlap fermion 
determinant. 

The reason for this procedure is that
direct simulations of overlap fermions using the HMC algorithm are 
very difficult and face the generic problem that the kernel operator 
used in the sign function of the overlap operator can develop 
very small eigenvalues. 
We have tried to avoid this problem by replacing $D_\mathrm{ov}$ 
by some operator $D_\mathrm{ov}^\mathrm{approx}$ that is a good 
approximation to $D_\mathrm{ov}$ but which is safe against these
low-lying eigenvalues. 

The general idea is to write 
%(we use for simplicity only a single 
%operator here, in practice one would have to use the operator 
%$D^\dagger D$, of course)
\be
\mathrm{det}D_\mathrm{ov}=\mathrm{det}D_\mathrm{ov}^\mathrm{approx}\cdot
\mathrm{det}\left[\frac{D_\mathrm{ov}}{D_\mathrm{ov}^\mathrm{approx}}
\right]
\equiv \mathrm{det}D_\mathrm{ov}^\mathrm{approx}\cdot \R\; .
\ee
One way to implement this idea is to use solely
$D_\mathrm{ov}^\mathrm{approx}$ 
for the HMC algorithm of the gauge fields. 
The remaining correction determinant ratio $\R$ could be 
implemented as a reweighting factor in the computation of a given
observable. 
Another possibility is to use $D_\mathrm{ov}^\mathrm{approx}$
as the guidance Hamiltonian in the molecular dynamics part of the HMC
algorithm only while the exact overlap operator $D_\mathrm{ov}$ is
used to evaluate the Hamiltonian for the accept/reject step.

A crucial question in the first of these approaches is, whether an operator 
$D_\mathrm{ov}^\mathrm{approx}$ can  be found such that the 
fluctuations in $\R$ are small enough to obtain statistically 
significant results. 
We decided therefore to test 
this idea by computing $\R$ on a number of gauge field configurations 
generated in the pure gauge theory at $\beta=3$ on a $L=16$ lattice. 

As approximations to the overlap operator we used 
the hypercube operator $D_\mathrm{hyp}$, an overlap operator with a
modified Chebyshev polynomial to evaluate the sign-function and an  
explicitly infra-red regulated operator, where 
\be
D_0/\sqrt{D_0^\dagger D_0} \rightarrow D_0/(\sqrt{D_0^\dagger D_0+\delta)}\; ,
\label{signmod}
\ee
with $D_0$ representing the kernel operator to construct the overlap 
operator. 
Out of these set of operators, it turned out that only the operator 
given in eq.~(\ref{signmod}) led to an acceptable size of
the fluctuations for the 
determinant ratio $R_\mathrm{det}$. However, even in this case, when 
the value of $\delta$ is chosen to be comparable to values of the 
low-lying eigenvalues of the kernel of the overlap operator, 
the fluctuations in 
$R_\mathrm{det}$ became so large that this operator does not seem to 
be practical when the reweighting technique is used, see 
ref.~\cite{schwingerpap} for a more detailed discussion. 

For the hypercube operator it turned out that although the mass parameter
was tuned in a wide range, $R_\mathrm{det}$ showed always large
fluctuations such that no stable result for a physical observable 
could be obtained. There the square root of the kernel is computed by a
Chebyshev polynomial $P_{n,\varepsilon}$  of the degree $n$ in the
interval $[\varepsilon,1]$. In the studies of the modified Chebyshev polynomial
$P_{\tilde n,\tilde \varepsilon}$ we found that it is possible to decrease the degree
$n$ of the polynomial to $\tilde n < n$, while keeping
$\tilde\varepsilon=\varepsilon$ , by about a factor of three. Although
this certainly accelerates the simulations, we also found that already
a slight increase of the lower bound $\epsilon$ of the polynomial
approximation to $\tilde\varepsilon > \varepsilon$ led again to large
fluctuations of $R_\mathrm{det}$.   

%Although from the above discussion we had to conclude that for the 
%approximate overlap operators we have tested here, the reweighting with
%$R_\mathrm{det}$ does not lead
%to stable results and rather showed large and UN-acceptable fluctuations. 
%We note that, of course, it would be possible
%to further improve the behaviour of $R_\mathrm{det}$ by using smearing 
%techniques. However, we did not follow this path any further. 

\begin{figure}[tbp]
\centering
\includegraphics[width=11.0cm]{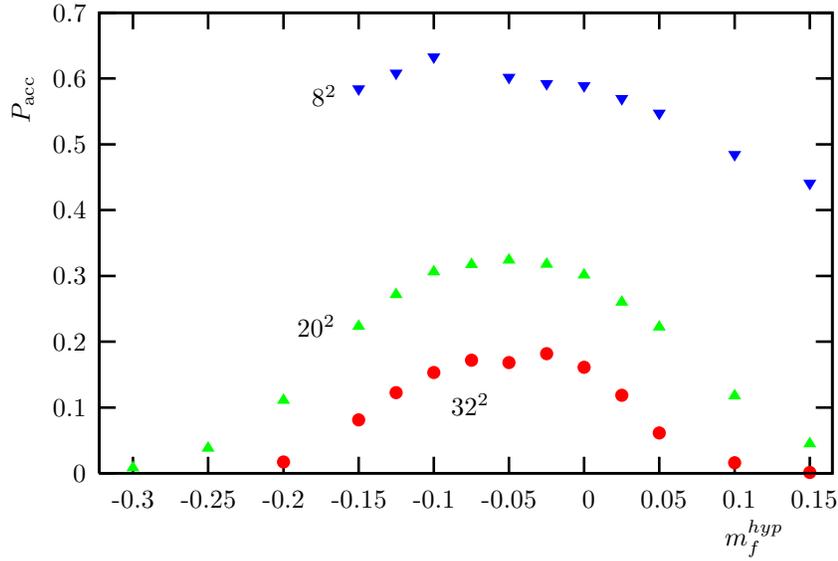}
  \caption{The acceptance rate $P_\mathrm{acc}$
using the hypercube operator as
the guidance Hamiltonian in the Hybrid Monte Carlo algorithm.
$P_\mathrm{acc}$ is plotted as a function of the bare overlap
fermion mass $m_f^\mathrm{hyp}$ for three lattice volumes.}
\label{fig:accept}
\end{figure}

As a second attempt, we
used an approximate overlap operator for the 
molecular dynamics part of the HMC algorithm.  
The setup is then to use 
$D_\mathrm{ov}$ for the accept/reject Hamiltonian while 
$D_\mathrm{ov}^\mathrm{approx}$ will be used only for the guidance 
Hamiltonian. 
We started to investigate this idea by using the hypercube 
operator as the guidance Hamiltonian. In fig.~\ref{fig:accept} we show
the acceptance rate as a function of the bare hypercube fermion mass parameter
$m_f^\mathrm{hyp}$ 
on three lattice volumes. 
The results are very surprising. 
From our previous experience with computing $R_\mathrm{det}$ we expected 
a quite bad behaviour of the acceptance rate $P_\mathrm{acc}$. 
In contrast, we find that 
$P_\mathrm{acc}$ is rather large, at least for the $8^2$ lattice. Even on 
our largest lattice ($32^2$) the acceptance rate does not go to zero but 
assumes reasonable values. Fig.~\ref{fig:accept} also demonstrates that 
$P_\mathrm{acc}$ develops a maximum at a certain value of the bare 
hypercube fermion mass that does not vary strongly with the lattice size.
We remark that for a certain threshold step size 
we could not increase the acceptance rate by decreasing the 
step size keeping the trajectory length fixed. 
This means that at this point we are sampling only configurations that 
are distributed according to the hypercube action.

%\begin{figure}[tbp]
%  \centering
%\includegraphics[height=10.0cm,width=12.0cm]{plots/accVsBareWith8x8.eps}
%  \caption{The acceptance rate $P_\mathrm{acc}$ 
%using the hypercube operator as 
%the guidance Hamiltonian in the Hybrid Monte Carlo algorithm. 
%$P_\mathrm{acc}$ is plotted as a function of the bare overlap 
%fermion mass at three lattice volumes.}
%  \label{fig:accept}
%\end{figure}

\section{Conclusions}

In this contribution we have tested four different lattice fermions 
in their approach to the continuum limit in the 2-dimensional massive
Schwinger model with $N_f=2$ flavour of dynamical fermions. At fixed
scaling variable $z= (m_f\sqrt{\beta})^{2/3}=0.2,0.4,0.8$ we have 
computed the pseudo scalar mass $m_\pi\sqrt{\beta}$ 
for various values of $\beta=1/e^2a^2$. 

For all kind of fermions used, Wilson, hypercube, maximally twisted 
mass and overlap fermions, the scaling behaviour of $m_\pi\sqrt{\beta}$ 
appears to be linear 
in $a^2$ and lead to consistent continuum values, demonstrating nicely
universality of the results. 
The precision of our simulation results reveal that the analytical 
calculations do not provide a satisfactory description of the data 
for all values of $z$.

We also presented possibilities to simulate overlap fermions dynamically
by replacing the exact overlap operator by an approximate version which
is used either as only the guidance Hamiltonian or as the full Hamiltonian
in the Hybrid Monte Carlo algorithm.
In the latter case we tested whether the approximation error can be
corrected by computing the determinant ratio of the exact and approximate
overlap operators stochastically.
This way turned out to be not successful since this (correcting) determinant
ratio showed large and un-acceptable fluctuations.

Changing the guidance Hamiltonian showed promising results, however.
Using the hypercube operator, which led to large fluctuations in  
evaluating the determinant ratio $R_\mathrm{det}$, as the guidance 
Hamiltonian we find reasonable acceptance rates. Unfortunately, 
$P_\mathrm{acc}$ drops significantly when the lattice volume is increased. 
If we think of simulations in four dimensions, we therefore have to 
conclude that the hypercube operator will in this case not be a good 
choice for the guidance Hamiltonian. 

Nevertheless, given the fact that the other approximate overlap 
operators, modified Chebyshev
polynomial and the operator in eq.~(\ref{signmod}), led to much more stable 
results for the fluctuations of $R_\mathrm{det}$, we believe that using 
these operators as the guidance Hamiltonian should lead to much 
improved acceptance rates. We are presently testing this idea.

\section{Acknowledgements}
We thank W. Bietenholz, V. Linke, C.~Urbach, U.~Wenger 
for many useful discussions. 
In particular we thank C.B. Lang for pointing out to us to use the 
modified sign function of eq.~(\ref{signmod}). 
The computer centers at DESY, Zeuthen, and at the Freie Universit\"at 
in Berlin provided 
the necessary technical help and computer
resources. 
%This work was supported by the DFG 
%Sonderforschungsbereich/Transregio SFB/TR9-03.

% see also: /usr/local/teTeX/share/texmf/bibtex/bst
\bibliographystyle{JHEP-2}
\bibliography{schwingerproc}

\end{document}